# Blue Organic Light-Emitting Diodes with External Quantum Efficiencies over 20% Based on Europium(II) Emitters

*Mahmoud Soleimani,* [a,b] *Toni Bärschneider,* [b] *Felix Kaden,* [b] *Roman Tkachov,* [b] *Sebastian Schellhammer,* [a] *Sebastian Reineke,*[a] *Carsten Rothe* [b]

[a] Institute of Applied Physics (IAP) and Dresden Integrated Center for Applied Physics and Photonic Materials (IAPP), Technische Universität Dresden, Dresden, Germany
[b] beeOLED GmbH, Dresden, Germany

## Abstract

The realization of blue electroluminescence with high efficiency and lifetime remains a long-standing hurdle for organic light-emitting diode (OLED) technology to overcome. Divalent Europium [Europium(II)] complexes offer a fundamentally distinct pathway toward this goal, as their atomic 4*f*–5*d* transitions yield single-Gaussian, spectrally pure emission with theoretical 100% exciton utilization in electroluminescence and no involvement of fragile organic bonds in the emissive process. Here, we present a rigid aza-crown europium(II) complex (Eu5NHCrown) that achieves near-unity photoluminescence quantum yield with bright, pure-blue emission and its incorporation in OLEDs. The emitter complex sublimes without decomposition and can be processed by industry-standard vacuum deposition. A bottom-emitting, single-host OLED architecture delivers an external quantum efficiency (EQE) of 20.7% with minimal roll-off (19.3% at $1000\ \mathrm{cd\,m^{-2}}$) and a narrowband electroluminescence with Commission Internationale de l'Éclairage (CIE) coordinates of (0.12, 0.25). This ligand design provides enhanced steric shielding of the Eu(II) center, enabling the highest reported efficiency among vacuum-deposited Eu(II)-based blue OLEDs while maintaining performance at high luminance. These results reveal the true potential of divalent Europium 4*f*–5*d* transitions for high-efficiency blue OLEDs, establishing a molecular design concept that bridges atomic-transition efficiency with the processability of organic materials.

## 1 Introduction

Organic light-emitting diode (OLED) displays are inseparable from modern life.[1] Yet despite their commercial success, OLEDs continue to face intrinsic challenges in achieving long operational lifetimes and high efficiencies—particularly for blue emission, which remains the bottleneck for full-color display technology.[2–4] Over the past decades, multiple emitter concepts have been explored, including phosphorescent, thermally activated delayed fluorescent (TADF), and conventional fluorescent materials.[5–10] Each material class has met important key performance parameters like color quality, lifetime, or efficiency, but none has reached industry-level performance in all parameters simultaneously. Phosphorescent emitters offer high exciton utilization yet suffer from instability due to bond cleavage, while fluorescent emitters exhibit excellent stability but are intrinsically limited in their ability to harvest triplet excitons.[2–4] These issues are mainly rooted in molecular orbital transitions, which inherently couple emissive states to chemical bonds.

Lanthanide-based emitters offer a fundamentally different strategy. Their emission can originate from metal-centered atomic $5d$–$4f$ transitions that are decoupled from molecular bonding, eliminating bond cleavage as a degradation pathway and enabling full exciton utilization independent of spin statistics. Among these, Eu(II) stands out as a particularly promising candidate due to its zero orbital angular momentum ($L = 0$) ground state, which yields a single-peak emission band and intrinsically short excited-state characteristics with lifetimes in the range of hundreds of nanoseconds to approximately 1 μs, characteristic of a parity-allowed $5d \rightarrow 4f$ transition.[11–13] Despite the conceptual appeal of Eu(II) emitters, progress toward highly efficient blue OLEDs has remained challenging. The realization of high-performance blue OLEDs would also open up future development of white OLEDs again, as they similarly lack a robust and efficient way to generate blue photons.[14]

Over the past 30 years, numerous studies have explored Eu-based emitters to demonstrate the above-mentioned theoretical advantage in the device. One of the earliest reports, Shipley et al. in 1999, described an orange-emitting Eu(II) borate complex, though with poor electroluminescent performance.[15] In 2018, Allen and co-workers reported a Eu(II) azacryptand complex with methyl substitution at the NH positions exhibited narrow, deep-blue emission (450 nm) and photoluminescence quantum yield (PLQY, $\Phi_{PL}$) of ≈ 100%. However, the resulting complex demonstrated salt-like character and poor volatility, rendering it unsuitable for vacuum deposition OLEDs.[12] Later in 2020, Liu *et al*. reported a related azacryptand complex that enabled OLED fabrication via vacuum deposition, achieving a high external quantum efficiency (EQE) of 17.7% and confirming that Eu(II) complexes can effectively harvest triplet excitons.[13] However, in this case, the specific coordination environment of Eu(II) led to a broad emission spectrum with a double-band feature in the green–yellow region, rather than the narrow blue Gaussian profile characteristic of $4f$–$5d$ transitions.[13] More recently, in 2024, Liu *et al*. reported a series of Eu(II) complexes employing bulky tert-butylated tris(pyrazolyl)borate ligands, achieving improved emission color and sublimation properties. The best-performing device exhibited sky-blue electroluminescence with a peak wavelength ($\lambda_{EL}$) of 478 nm, a full width at half maximum (FWHM) of 64 nm, Commission Internationale de L'Eclairage (CIE) coordinates of (0.13, 0.27) at 1000 cd $m^{-2}$, and a maximum EQE of 15.7%.[16] In our previous work, we introduced two new concepts for Europium-based emitters incorporating a crown ether and carborate anions. We showed that this combination provides an effective strategy to stabilize the Eu(II) core while maintaining strong $4f$–$5d$ emission, volatility, and deep blue emission.[17] Beyond these advances, the realization of a vacuum-processable Eu(II) emitter enabling a highly efficient blue OLED with near-limit theoretical efficiency while maintaining its efficiency at practical luminance had not yet been achieved.

Building on the foundation of our previous work, we introduce here Eu5NHCrown, a Eu(II) complex with an advanced crown-ether ligand design that stabilizes the divalent center, preserves its atomic-transition emission characteristics, and provides device-relevant energy-level alignment favorable for efficient electroluminescence. The design enables decomposition-free sublimation and near-unity PLQY. OLEDs incorporating Eu5NHCrown exhibit bright blue emission with an electroluminescence peak wavelength of 480 nm, a full width at half maximum of 44 nm corresponding to CIE (x, y) coordinates of (0.12, 0.25) with a maximum EQE of 20.7% that only slightly decreases to 19.3% at 1000 $cd\,m^{-2}$. These results establish Eu5NHCrown as the first europium-based emitter to approach the theoretical efficiency limit in OLEDs, delivering the highest reported efficiency for a blue Eu(II)-based OLED and setting a new benchmark for blue divalent europium devices based on atomic-transition emitters.

# 2 Results and Discussion

## 2.1 Design and Synthesis

Efficient Eu(II)-based blue electroluminescence requires a combination of blue 4*f*–5*d* emission, steric shielding of the Eu(II) center, suitable energy-level alignment, and vacuum processability. To meet these requirements, we designed Eu5NHCrown, a Eu(II)-based emitter featuring an aliphatic all-aza-20-crown-5 ether as the neutral ligand and two carborate anions. The different size and type of this neutral ligand are the main differences from the previously introduced Eu-based emitters EuCrown and EuCovCrown.[17] Specifically, instead of an 18c6-type crown with two nitrogen and four oxygen donors, Eu5NHCrown comprises a larger 20-ring with five nitrogen donors connected through propylene bridges, while it uses the same carborate $[CB_{11}H_{12}]^{-}$ anions to neutralize divalent europium. The differences in the ring size and donor groups with respect to our previous compounds have two important consequences: Firstly, being larger than the 18-ring, the 20-ring folds in on itself to accommodate the Europium(II) cation, which improves the steric shielding of the Europium(II) center (*cf.* Figure 1A). This folding effect is further enhanced by the more flexible propyl-linking groups and the kinks introduced at the triangular [–Pr–**N**(H)–Pr–] nitrogen donors, which is in contrast to the ethylene-linked and mostly linear [–Et–**O**–Et–] oxygen donors of the 18c6 crowns. Secondly, polarized through the coordination of Eu(II), the NH groups help to bind the carborate anions via dihydrogen bonds[18], which is evident from four short ($<$ 2.2 Å) non-covalent H–H contacts in the structure displayed in Figure 1A as dotted red lines. With this new approach, we combine the good steric shielding of EuCovCrown with the deep energy levels of EuCrown.

The calculated electronic parameters for Eu5NHCrown obtained at the TD-DFT/ωB97X-D3/SMD/LR-PCM level of theory with ORCA 6.0.1[19,20] for the lowest energy conformer provided by the CREST/CENSO workflow [21–23] which has recently been demonstrated to work particularly well for Eu [24] are summarized in Figure 1, including an emission energy ($E_{ems}$) of 2.62 eV, an ionization energy (IE) of 5.88 eV, an excited-state ionization energy (ES-IE) of 2.88 eV, and the corresponding natural transition orbitals (NTOs).

The complex Eu5NHCrown was synthesized by reacting equimolar amounts of $\mathbf{Eu\,(CB_{11}H_{12})_2{\cdot}5THF}$ and the macrocyclic amine $\mathbf{5NH_5Pr}$ in toluene under inert conditions, affording a colorless powder in high yield (89%). The product was purified by sublimation at $10^{-6}$ mbar to give an analytically pure material (yield of 87%) suitable for vacuum deposition. Detailed experimental procedures and characterization data are provided in the Supporting Information.

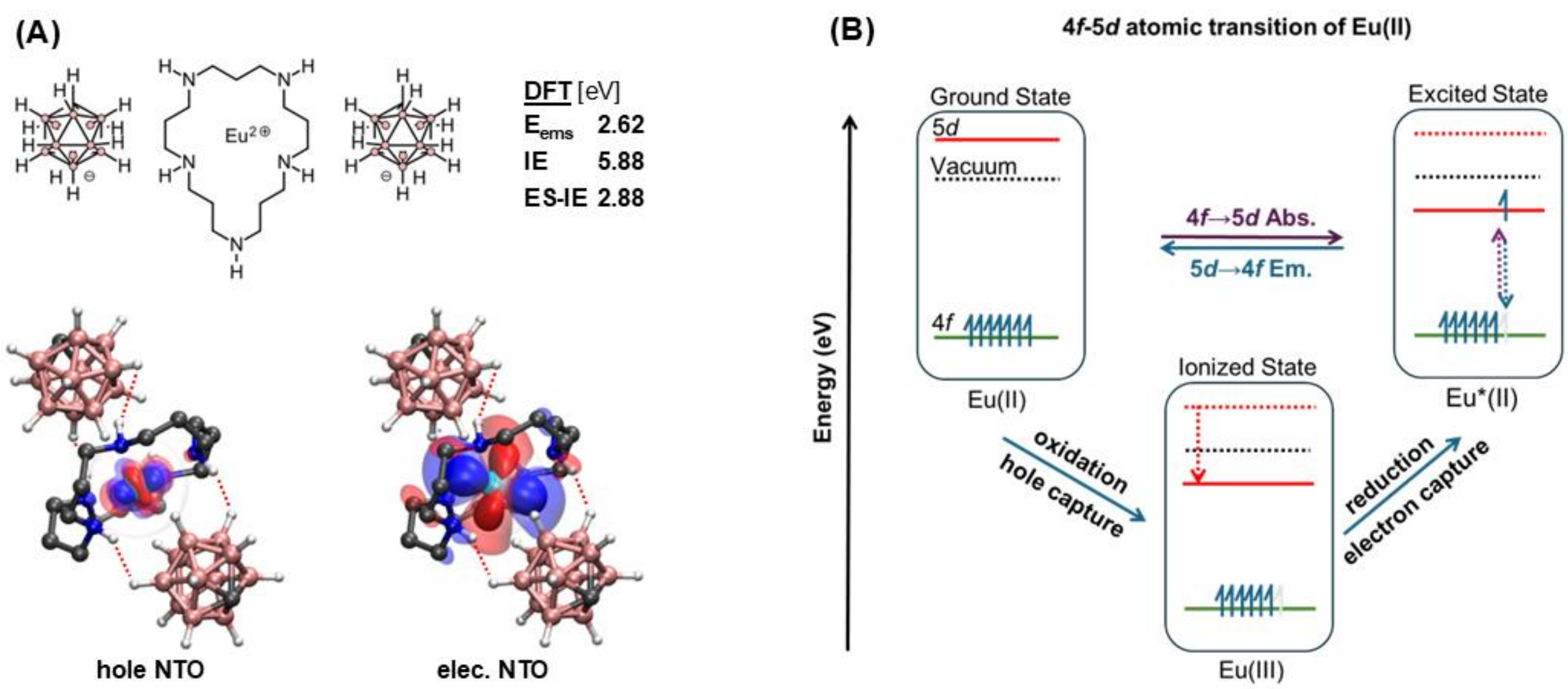

**Figure 1. (A)** 2D and 3D structures of Eu5NHCrown, along with emission energies and hole and electron natural transition orbitals (NTOs) of the lowest lying excited state at its relaxed structure, adiabatic ionization energy (IE), and excited-state ionization energy (ES-IE), all calculated at the TD-DFT/$\omega$B97X-D3/SMD/LR-PCM level of theory. Shown is the first NTO-pair ($4f$-hole on the left, $5d$-electron on the right), which makes up $\approx$ 99.5% of the transition. The four dihydrogen bonds are indicated by dotted red lines. Visualization using VMD 1.9.2[25] with isosurfaces for 0.94e (solid) and 0.98e (transparent). For clarity, all hydrogen atoms bonded to carbon are hidden. **(B)** Schematic energy-level diagram illustrating the atomic $4f$–$5d$ transition mechanism of Eu(II). In the Eu(II) ground state, the $5d$ level lies above the vacuum level and is therefore not bound. Upon optical excitation, absorption promotes an electron from the $4f$ shell to the $5d$ state, followed by radiative $5d \rightarrow 4f$ emission. In electroluminescence, hole capture (injection) first oxidizes Eu(II) to an Eu(III)-like ionized state, in which the $5d$ level becomes stabilized below the vacuum level, enabling electron capture (injection) into the bound $5d$ state to form Eu*(II). The excited Eu*(II) state subsequently emits through the characteristic atomic $5d \rightarrow 4f$ transition.

## 2.2 Photophysics and Thermal Properties

The photophysical and thermal properties of Eu5NHCrown are systematically characterized in degassed toluene solution and under vacuum, respectively. Steady-state photoluminescence (PL) and excitation spectra, time-resolved photoluminescence decay, and thermogravimetric (TGA) analysis are shown in Figure 2, while the key photophysical parameters are summarized in Table 1. The photoluminescence spectrum of Eu5NHCrown exhibits an intense single-peak blue emission centered at 482 nm ($\approx$ 2.57 eV) with a full width at half maximum of 58 nm in toluene solution. Time-resolved measurements reveal a monoexponential decay with a lifetime of approximately 1 μs, which together with the single peak emission profile, is consistent with the $4f^7$ ($^8S_{7/2}, L = 0$) ground state of Eu(II) and a spin-allowed $5d \rightarrow 4f$ transition, confirming that the emission originates from the Eu(II) center rather than from ligand or charge-transfer states. Compared with previously reported Eu(II) complexes such as EuCrown and EuCovCrown, Eu5NHCrown exhibits a red-shifted emission by $\approx$0.15 eV while maintaining spectral purity, reflecting the influence of the stronger ligand field and the more rigid coordination environment provided by the ligand for Eu(II).

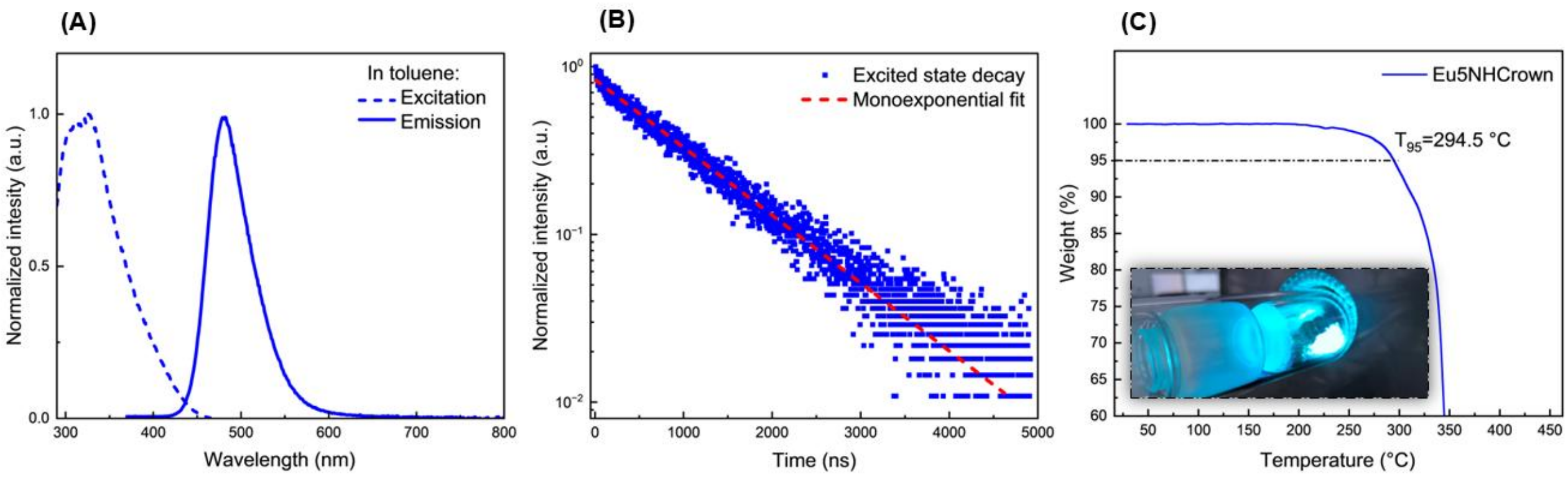

Figure 2. Photophysical properties and thermal stability of Eu5NHCrown. **(A)** Normalized emission and excitation spectra of Eu5NHCrown in degassed toluene solution with a concentration of $10^{-5}$ M; the corresponding photoluminescence parameters are summarized in Table 1. **(B)** Time-resolved photoluminescence decay of Eu5NHCrown on a semi-logarithmic scale in degassed toluene solution with a concentration of $10^{-5}$ M. The sample was excited at 300 nm, and the monoexponential fit yields the excited-state lifetime $\tau$ reported in Table 1. **(C)** Thermogravimetric analysis (TGA) curve of Eu5NHCrown, showing a 5% weight-loss temperature of $T_{95}$ = 294.5 °C. The inset shows Eu5NHCrown during the sublimation experiment under ultraviolet (UV) illumination, highlighting its bright blue emission and suitability for vacuum processing.

Table 1. Photophysical data for Eu5NHCrown in degassed toluene with a concentration of $10^{-5}$M.

| **Emitter** | $\lambda_{\mathrm{em}}$ (nm) | FWHM (nm) | $\Phi_{\mathrm{PL}}$ (%) | $\tau$ (ns) | $k_{\mathrm{r}}$ ($\mathrm{s}^{-1}$) | $k_{\mathrm{nr}}$ ($\mathrm{s}^{-1}$) |
|---|---|---|---|---|---|---|
| Eu5NHCrown | 482 | 58 | 95 | 1078 | $88.1 \times 10^4$ | $4.6 \times 10^4$ |

$\lambda_{\mathrm{em}}$: peak emission wavelength; FWHM: full width at half maximum; $\Phi_{\mathrm{PL}}$: photoluminescence quantum yield; $\tau$: excited-state lifetime; $k_{\mathrm{r}}$ and $k_{\mathrm{nr}}$ are the radiative and non-radiative rate constants, respectively, calculated using $k_{\mathrm{r}} = \Phi_{\mathrm{PL}}/\tau$ and $k_{\mathrm{nr}} = (1 - \Phi_{\mathrm{PL}})/\tau$, assuming that all absorbed photons reach the emitting state.

The photoluminescence quantum yield of Eu5NHCrown reaches 95% in toluene, demonstrating highly efficient emission with strongly suppressed nonradiative decay. Such high efficiency places this complex among the best Eu(II) emitters reported to date, comparable to the near-unity yields of rigid azacryptate and pyrazolylborate systems.[11] Using the measured photoluminescence quantum yield ($\Phi_{\mathrm{PL}}$ = 95%) and excited-state lifetime ($\tau$ = 1078 ns), the radiative and nonradiative rate constants are estimated as $88.1 \times 10^4$ $\mathrm{s}^{-1}$ and $4.6 \times 10^4$ $\mathrm{s}^{-1}$, respectively. The dominance of radiative decay supports an efficient Eu(II)-centered 5*d*–4*f* emission process with limited nonradiative relaxation. The slightly longer emission lifetime compared with some reported Eu(II) emitters likely reflects subtle differences in ligand-field strength and coordination rigidity, while the high photoluminescence quantum yield indicates that radiative decay remains highly efficient. This performance arises from the optimized ligand field of 20-crown-5 architecture, which confines the Eu(II) 5*d* orbital while minimizing vibrational coupling and quenching pathways. This structural property provides excellent steric shielding of the Eu(II) core and effectively localizes the excited state and restricts intramolecular relaxation, leading to high-efficiency and well-isolated luminescence – the latter beneficial for low OLED efficiency roll-off.

Thermogravimetric analysis under vacuum confirms the thermal stability and volatility of Eu5NHCrown (Figure 2C). The onset of volatilization occurs at 255.3 °C for 1% weight loss, in good agreement with the experimental sublimation temperature (260 °C), and the 5% weight-loss temperature ($T_{95}$) corresponds to 294.5 °C. A photograph of Eu5NHCrown during the sublimation experiment under ultraviolet (UV) illumination is shown in the inset of Figure 2C. Further experimental details are provided in the Supporting Information. These results confirm the suitability of Eu5NHCrown for vacuum processing and vacuum-deposited OLEDs. Combined with its near-unity photoluminescence quantum yield and blue emission, this thermal and processing stability suggests that Eu5NHCrown is a promising emitter for efficient vacuum-deposited devices, as demonstrated in the following section.

### 2.3 Vacuum-fabricated OLEDs

To evaluate the electroluminescent performance of Eu5NHCrown, bottom-emissive OLEDs were fabricated using a conventional multilayer architecture: ITO (Indium tin oxide) / $MoO_3$ (Molybdenum trioxide) [2 nm] / SimCP2 (Bis[3,5-di(9H -carbazol-9-yl)phenyl]diphenylsilane) [70 nm] / SiDBFCz (9-[4-(Triphenylsilyl)-2-dibenzofuranyl]-9H-carbazole):Eu5NHCrown [30 wt%, 20 nm] / mSiTrz (2-Phenyl-4,6-bis(3-(triphenylsilyl)phenyl)-1,3,5-triazine) [5 nm] / TSPO1 (Diphenyl[4-(triphenylsilyl)phenyl]phosphine oxide):Yb (Ytterbium) [1 vol%, 25 nm] / Al [120 nm]. SiDBFCz was selected as the host to satisfy the key requirements for efficient Eu5NHCrown electroluminescence: a wide bandgap with high triplet energy for exciton confinement, a deep HOMO for hole trapping on the emitter, suitable LUMO alignment for electron confinement, and electron-transport capability to enable charge balance. This simplified configuration was chosen to highlight the intrinsic emissive efficiency of the Eu(II) complex via atomic transition without the aid of co-hosts or exciplex-forming layers. An $Ir(ppy)_3$-based control device was fabricated using a similar layer stack as the Eu5NHCrown-based blue-emitting device, except that the thickness of the electron-transport layer (ETL) was adjusted to match the optical cavity to the green emission of $Ir(ppy)_3$. Details on device optimization, pixel statistics, and control devices are provided in the Supporting Information.

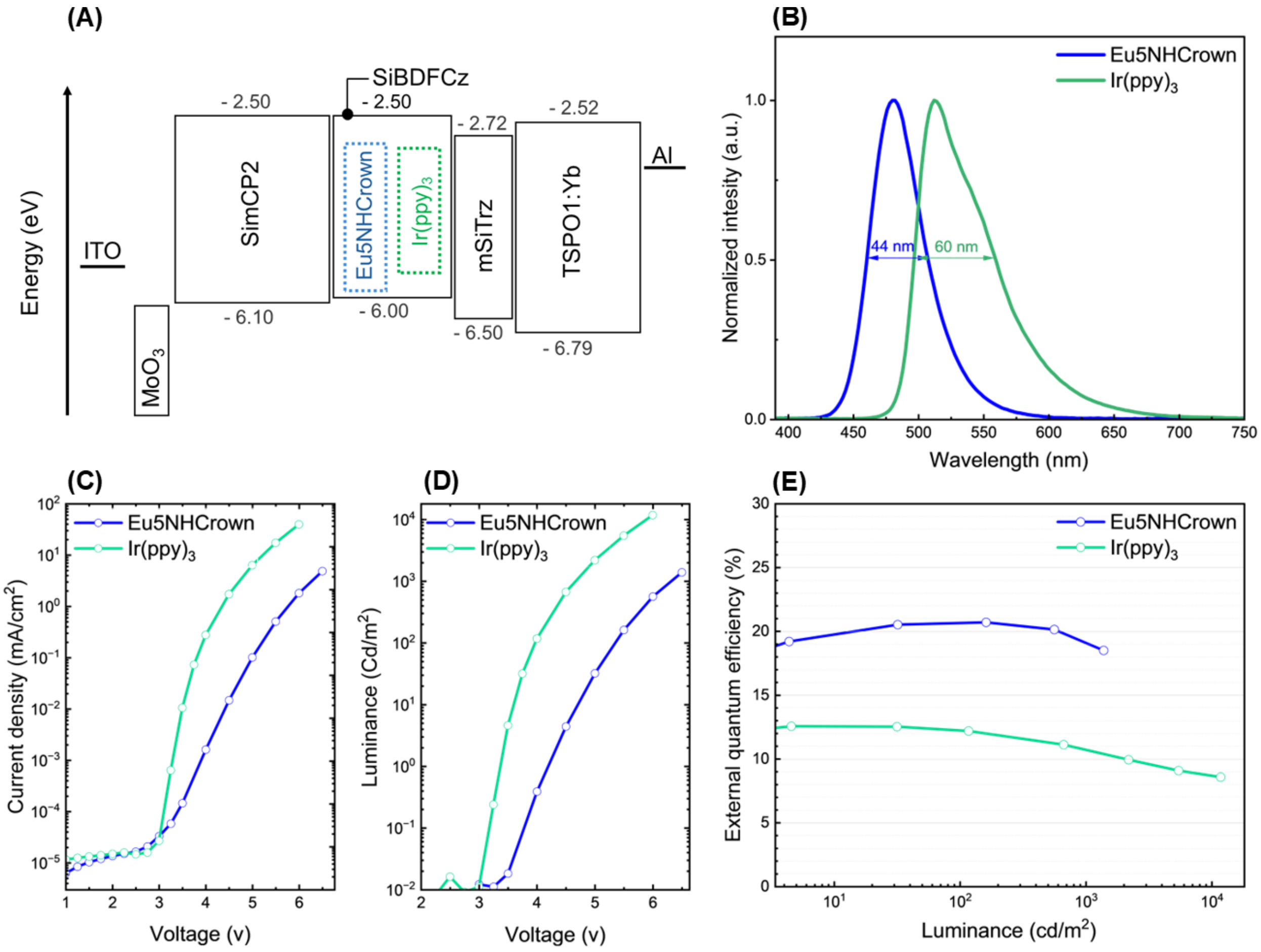

Figure 3. Electroluminescent performance and comparison of Eu5NHCrown- and Ir(ppy)3-based OLEDs. **(A)** Energy-level diagram of the bottom-emitting OLED architecture, showing Eu5NHCrown and Ir(ppy)3 incorporated in the emissive layer. **(B)** Normalized electroluminescence spectra of the Eu5NHCrown-based OLED and Ir(ppy)3 control device, showing peak emissions ($\lambda_{EL}$) at 480 and 512 nm with full width at half maximum (FWHM) values of 44 and 60 nm, respectively. **(C)** Current density–voltage characteristics and **(D)** luminance–voltage characteristics of the devices. **(E)** External quantum efficiency (EQE) as a function of luminance. The Eu5NHCrown-based OLED shows a turn-on voltage of 4.2 V, a maximum EQE of 20.7%, and an EQE of 19.3% at 1000 cd m$^{-2}$; the detailed electroluminescent parameters are summarized in Table 2.

Table 2. Electroluminescent performance of the Eu5NHCrown-based OLED.

| **Emitter** | $V_{on}$ (V) | $EQE_{max}$ (%) | $EQE_{1000}$ (%) | $\lambda_{EL}$ (nm) | FWHM (nm) | CIE (x, y) |
|---|---|---|---|---|---|---|
| Eu5NHCrown | 4.2 | 20.7 | 19.3 | 480 | 44 | 0.12, 0.25 |

$V_{on}$: turn-on voltage or voltage required to reach a luminance of 1 cd m$^{-2}$; $EQE_{max}$: maximum external quantum efficiency; $EQE_{1000}$ external quantum efficiency at 1000 cd m$^{-2}$ ; $\lambda_{EL}$: peak electroluminescence wavelength; CIE (x, y): 1931 color coordinates at 1000 cd m$^{-2}$; FWHM: full width at half maximum at 1000 cd m$^{-2}$.

The energy-level diagram, normalized electroluminescence spectra, current density–voltage (*J–V*), luminance–voltage (*L–V*), and external quantum efficiency (EQE) characteristics of Eu5NHCrown and the Ir(ppy)3-based control device are presented in Figure 3, while the key performance parameters of the Eu5NHCrown-based OLED are summarized in Table 2. The Eu5NHCrown OLED achieves a maximum EQE ($EQE_{max}$) of 20.7% with a turn-on voltage of 4.2 V (defined at a luminance of 1 cd m$^{-2}$ ), representing a new performance benchmark for blue Eu(II)-based OLEDs; a comparison with previously reported Eu-based OLEDs is provided in the Supporting Information. No detectable host emission was observed in the electroluminescence spectra across the investigated operating voltage and luminance ranges, indicating that electroluminescence originates predominantly from Eu5NHCrown emitters. The Eu5NHCrown-based OLED exhibits a blue emission with an electroluminescence peak wavelength of 480 nm, a full width at half maximum of 44 nm, and CIE (x, y) coordinates of (0.12, 0.25) at 1000 cd m$^{-2}$. As illustrated in Figure 1B, Eu(II) does not possess a bound $5d$ state in the electronic ground state, as the $5d$ manifold lies above the vacuum level. Therefore, electroluminescence first requires oxidation of the complex by hole capture, or equivalently, extraction of one electron from the half-filled $4f$ shell, forming a transient ionized Eu(III)-like state. In this ionized state, the atomic orbitals contract and the $5d$ level is stabilized below the vacuum level, allowing electron capture into the bound $5d$ state to form Eu*(II), which then relaxes radiatively through the characteristic $5d$→$4f$ transition. Since Eu(II) emission arises from an atomic orbital $5d$–$4f$ transition rather than molecular orbitals, the transition dipoles are inherently isotropic. The emission terminates on the $4f^7$ ($^8S_{7/2}$) ground state, which possesses eight energetically equivalent spin sublevels and zero orbital angular momentum ($L$ = 0), leading to a single-band emission with an isotropic orientation-independent emission.

Thus, considering the high photoluminescence quantum yield of Eu5NHCrown ($\Phi_{PL}$ of 95%) and the typical optical outcoupling efficiency of randomly oriented emitters in conventional bottom-emitting OLEDs ($\eta_{out}$ ≈ 20–22%), the measured EQE of 20.7% corresponds to near-unity exciton utilization efficiency (EUE). Assuming near-unity charge balance and using the upper end of this outcoupling range ($\eta_{out}$ ≈ 22%) the measured EQE corresponds to an exciton utilization efficiency of approximately 99%, indicating that the device operates close to the maximum EQE expected for 100% exciton utilization in this device architecture. The upper limit of EQE is primarily constrained by the theoretical maximum of the optical outcoupling efficiency.[16,26,27] A detailed analysis of this limit is provided in the Supporting Information. Remarkably, the device maintains an EQE of 19.3% at 1000 cd m$^{-2}$, corresponding to an efficiency roll-off of only 6.8% relative to the maximum EQE. This low roll-off indicates that efficiency-loss processes remain strongly suppressed at practical luminance, consistent with the spin and parity-allowed $5d$–$4f$ transition of Eu(II) which enables efficient exciton harvesting without the need for intersystem crossing or triplet-management processes.

## 3 Discussion and Outlook

Eu(II)-based emitters offer an attractive route toward efficient blue OLEDs because their atomic $5d$–$4f$ transitions can provide spin-allowed emission and full exciton utilization. However, the full potential of Eu(II) emitters, combining narrow blue emission, high photoluminescence efficiency, vacuum processability, and high electroluminescent efficiency in OLEDs at practical luminance within a single material platform, had not been demonstrated before. Eu5NHCrown represents a major advance among Eu(II)-based OLED emitters, combining blue emission, near-unity photoluminescence quantum yield (PLQY), clean sublimation and the highest reported device efficiency for a Eu(II)-based blue OLED. The optimized device reaches an external quantum efficiency (EQE) of 20.7% with low efficiency roll-off, maintaining an EQE of 19.3% at 1000 cd $m^{-2}$. This corresponds to only 6.8% loss relative to the maximum EQE and approaches the expected efficiency limit for isotropic emitters in a conventional bottom-emitting OLED architecture. This establishes Eu5NHCrown as a benchmark Eu(II)-based emitter that combines near-limit device efficiency with the photophysical and thermal properties required for vacuum-deposited OLEDs. Taken together, these results demonstrate that Eu5NHCrown bridges atomic-like transition efficiency and organic processability, establishing a new benchmark for blue OLEDs based on lanthanide emitters.

Beyond benchmarking against Eu-based systems, Eu5NHCrown also stands out in the broader context of blue OLED emitters. While many high-efficiency blue OLEDs rely on carefully optimized host combinations and exciton-management strategies to improve efficiency and reduce roll-off at high luminance, Eu5NHCrown achieves close-to-limit efficiency in a comparatively simple bottom-emitting architecture using a single host. This demonstrates that Eu(II) $5d$–$4f$ emitters can combine the efficient exciton utilization characteristic of inorganic atomic emitters with the vacuum processability and device compatibility of organic materials. Future development will focus on tuning the emission energy and advancing operational stability. Subtle modifications to the crown geometry and donor strength offer a rational handle to shift the $5d$–$4f$ transition toward deeper blue while maintaining high radiative efficiency. This strategy positions Eu(II) emitters as promising candidates with high potential for next-generation OLED display technology.

## 4 Experimental Section

**Materials**: All materials were obtained from commercial suppliers and further purified by resublimation prior to use. Other commercial materials or reagents were used as received unless otherwise noted. **Fabrication**: OLEDs were fabricated on pre-patterned and cleaned ITO glass substrates in a Kurt J. Lesker vacuum deposition system under high vacuum ($<1 \times 10^{-7}$ mbar). Organic and metal layers were sequentially deposited with precise rate control (0.5–1.0 Å $s^{-1}$ for organics, 0.2 Å $s^{-1}$ for $MoO_3$, and 2 Å $s^{-1}$ for Al). OLEDs were encapsulated inside a nitrogen glovebox using a glass cavity cap, UV-curable epoxy, and a getter. **Measurements**: Photoluminescence (PL) and photoluminescence quantum yield (PLQY) measurements were performed under an inert atmosphere using an Edinburgh Instruments spectrometer and a Hamamatsu Quantaurus-QY. Steady-state PL spectra were recorded at 300 nm excitation, and excited-state lifetimes were obtained by time-correlated single-photon counting (TCSPC) using 300 nm pulsed excitation. Emission was detected with a monochromator–photomultiplier setup at the Eu(II) peak wavelength (5–10 nm bandwidth) to selectively probe emitter emission. Current–voltage–luminance (*J*–*V*–*L*) characteristics and electroluminescence spectra were recorded using a calibrated McScience Inc IVL robot.

**Acknowledgments**
Financial support from the project "Blaue Emitter" (SAB no. 100671218 and 100671145) is gratefully acknowledged. The project is co-funded by the European Union and co-financed from tax revenues on the basis of the budget adopted by the Saxon State Parliament.

**Conflict of Interest**
M.S., T.B., R.T., F.K., and C.R. are employed at beeOLED GmbH, which develops lanthanide materials for OLED applications.  S.R. supports beeOLED GmbH as a scientific advisor. The mentioned authors hold options in beeOLED GmbH.

## Supporting Information

**Blue Organic Light-Emitting Diodes with External Quantum Efficiencies over 20% Based on Europium(II) Emitters**

*Mahmoud Soleimani,* [a,b] *Toni Bärschneider,* [b] *Felix Kaden,* [b] *Roman Tkachov,* [b] *Sebastian Schellhammer,* [a] *Sebastian Reineke,*[a] *Carsten Rothe* [b]

[a] Institute of Applied Physics (IAP) and Dresden Integrated Center for Applied Physics and Photonic Materials (IAPP), Technische Universität Dresden, Dresden, Germany
[b] beeOLED GmbH, Dresden, Germany

This Supporting Information contains experimental procedures, material characterization, and additional device data supporting the main manuscript. It includes details on the synthesis and purification of Eu5NHCrown, thermal stability and sublimation behavior, host characterization, OLED fabrication and optimization, pixel statistics, $Ir(ppy)_3$ control devices, theoretical EQE-limit analysis, and comparison with previously reported Eu-based OLEDs.

# 1. Preparation of the emitter

## 1.1 General Information

All reactions were performed under an inert nitrogen atmosphere in flame-dried glassware using standard schlenk techniques or a glove box. Tetrahydrofuran (THF, 99.5 %, water <50 ppm) and toluene (99.85%, water < 50 ppm) were purchased from Fisher Scientific and dried by passing through a column of molecular sieves followed by aluminum oxide using an MBraun Solvent Purification System. Trimethylammonium closo-1-carbadodecaborate (>97 %) was purchased from Katchem and used without further purification. 1,5,9,13,17-Pentaazacycloicosane (5NH5Pr) and Europium bis(bis(trimethylsilyl)amide) · 2 THF (Eu(HMDS)2) were prepared according to literature. All other reagents were used as purchased. NMR spectra were recorded on a Magritek Spinsolve 60. Chemical shifts ($\delta$) are reported in parts per million (ppm) downfield of tetramethylsilane. The complexes were characterized by mass spectrometry (MS) in solution (methanol), recorded on a Thermo Scientific MSQ Plus Mass Detector; elemental analysis (EA), performed on an Elementar UNICUBE, and photophysical measurements.

## 1.2 Preparation of Eu5NHCrown

**Europium bis(closo-1-carbadodecaborate) penta(tetrahydrofuran) ($Eu(CB_{11}H_{12})_2$ · 5 THF)**

A solution of 6.19 g (30.45 mmol, 2.0 eq.) trimethylammonium closo-1-carbadodecaborate in THF (80 mL) was added dropwise to a solution of 9.39 g (15.2 mmol, 1.0 eq.) $Eu(HMDS)_2$ in THF (100 mL). The resulting suspension was stirred for 1 h at room temperature and then filtered. The obtained solid was washed with THF and dried in a vacuum to yield 12.00 g (15.2 mmol, 99%) of the title compound as a white powder. This compound was used without further purification.

**Europium bis(closo-1-carbadodecaborate) 1,5,9,13,17-pentaazacycloicosane (Eu5NHCrown)**

A quantity of 4.02 g (5.03 mmol, 1.0 eq.) of $Eu(CB_{11}H_{12})_2$ was dissolved in hot toluene (200 mL). The synthetic route to Eu5NHCrown is shown in Figure S1. The hot solution was filtered through a syringe filter and added to a solution of 1.65 g (5.78 mmol, 1.0 eq.) 5NH5Pr in toluene (40 mL). The resulting mixture was stirred for 1 h at room temperature. The solvent was removed in vacuo. Hexane (40 mL) was added to the resulting oil, and the mixture was stirred for an additional 2 h at room temperature. The precipitate was collected by filtration, washed with hexane, and dried in a vacuum to yield 3.25 g (4.49 mmol, 89%) of Eu5NHCrown as a colorless powder. For further purification, the obtained powder was sublimed at 260 °C ($7 \times 10^{-6}$ mbar). Yield: 87 %. **MS (ESI):** $m/z$ = +286.4 $(5NH5Pr + H)^+$; -143.2 $(CB_{11}H_{12})^-$; -572.5 $(5NH5Pr + H + 2\,CB_{11}H_{12})^-$. **Elemental analysis:** Calcd.: C, 28.22; H, 8.22; N, 9.68. Found: C, 28.21; H, 7.97; N, 9.66. $\lambda_{max}$(emission) = 2.57 eV (toluene).

Figure S1. Synthesis of Eu5NHCrown. Reaction scheme showing the coordination of 1,5,9,13,17-pentaazacycloicosane (5NH5Pr) to $Eu^{2+}$, forming the Eu(II) complex Eu5NHCrown stabilized by two closo-1-carbadodecaborate $(CB11H12)^{-}$ anions.

## 2. Thermal stability

The thermal stability of Eu5NHCrown was investigated by thermogravimetric analysis (TGA, Figure S2). The material was measured in a 40 µl Al crucible without a lid under vacuum with a heating rate of 10 K/min and a typical sample size of 2–10 mg. The onset of volatilization, corresponding to 1% weight loss, was found at T = 255.3 °C, which is in accordance with the experimental sublimation temperature around T = 260 °C. The slow mass reduction up to T ≤ 330.5 °C can be mainly attributed to evaporation of the material, implying thermal stability to this point.

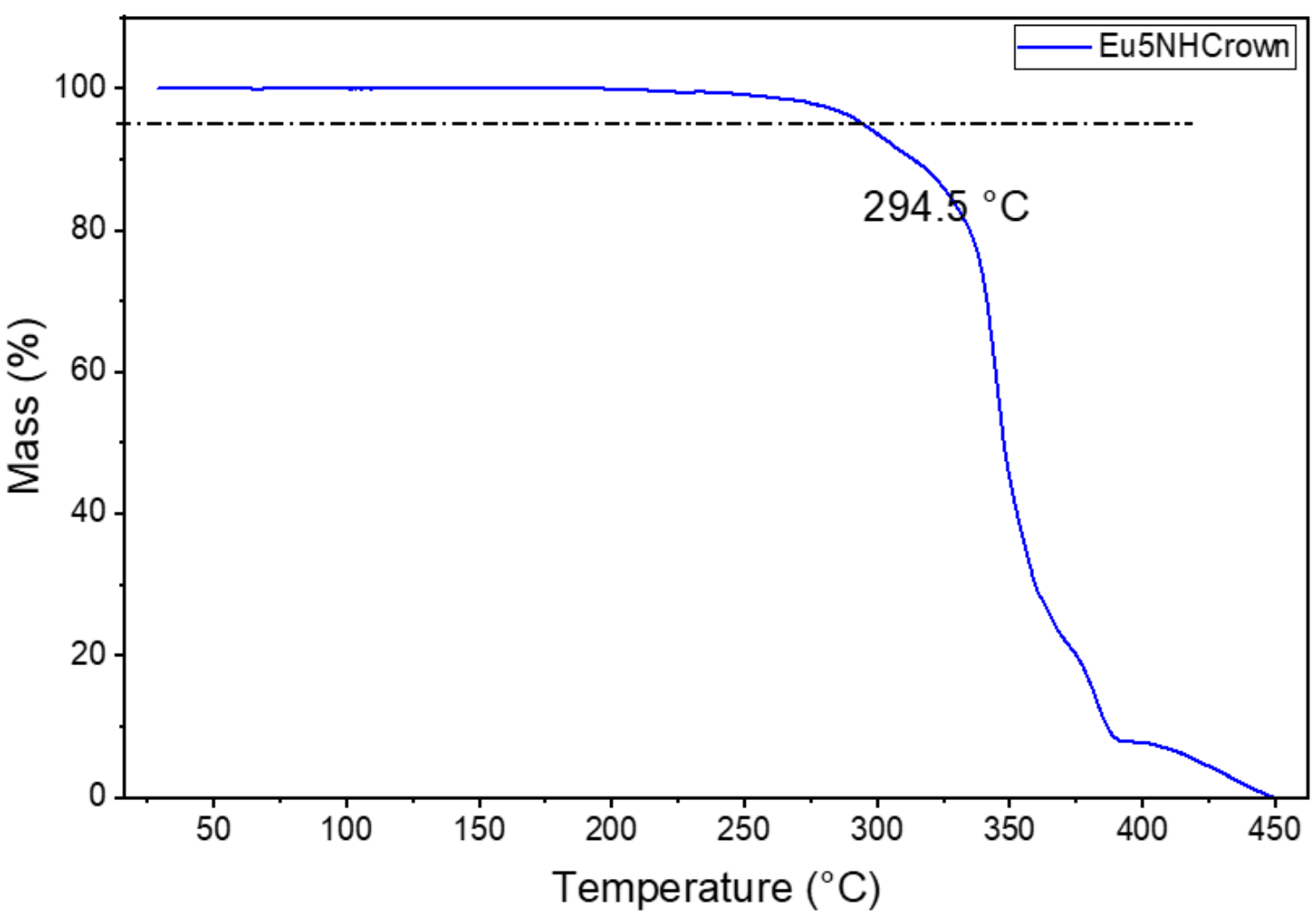

Figure S2. Thermogravimetric analysis of Eu5NHCrown. TGA curve of Eu5NHCrown measured under vacuum, showing a 5% weight-loss temperature of $T_{95}$ = 294.5 °C. The gradual mass loss before the main decomposition step is attributed to volatilization of the material, consistent with its sublimation behavior.

## 3. Organic Light-Emitting Diodes (OLEDs)

OLEDs were fabricated on pre-patterned ITO-coated glass substrates (sheet resistance ~ $15\ \Omega\ \mathrm{sq}^{-1}$), which were ultrasonically cleaned in detergent, deionized water, acetone, and isopropanol, followed by UV–ozone treatment to improve surface wettability. All organic and inorganic layers were deposited in a Kurt J. Lesker vacuum thermal-evaporation system operating at pressures below $1 \times 10^{-7}$ mbar. Deposition rates were precisely controlled using individual quartz-crystal microbalances (QCMs): 0.5–1.0 Å $\mathrm{s}^{-1}$ for organic layers, 0.2 Å $\mathrm{s}^{-1}$ for $MoO_3$, and 2 Å $\mathrm{s}^{-1}$ for Al. The layer thicknesses were monitored in real time during deposition and subsequently verified by a profilometer. Following deposition, devices were transferred directly into a nitrogen glovebox without exposure to ambient air and encapsulated using a glass cavity cap, UV-curable epoxy, and a moisture/oxygen getter to ensure operational stability during characterization. The chemical structures of the organic materials used in the device stack are shown in Figure S3.

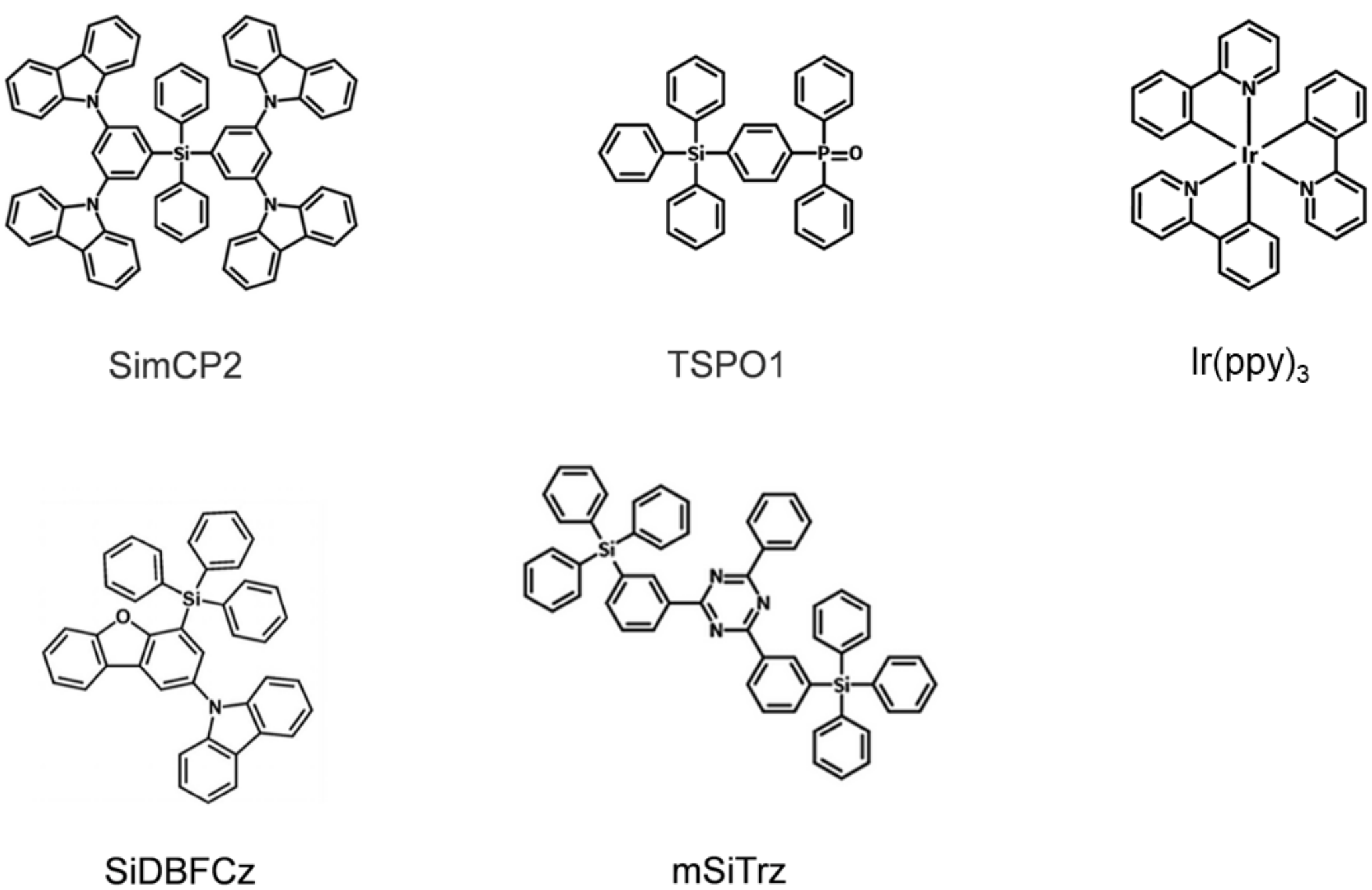

Figure S3. Chemical structures of organic materials used in the OLED device stack and control device. Chemical structures of SimCP2, TSPO1, SiDBFCz, mSiTrz, and $Ir(ppy)_3$ used for the fabrication of the Eu5NHCrown-based OLEDs and $Ir(ppy)_3$ control devices.

### 3.1. Organic host characterization

SiDBFCz was used as the host material in the emissive layer of the Eu5NHCrown-based OLEDs. This host was selected because it combines the key requirements for efficient blue electroluminescence: ambipolar charge-transport character to support balanced recombination, a wide optical gap and high triplet energy for exciton confinement, a deep HOMO level that promotes hole trapping on the emitter, and suitable LUMO alignment for electron confinement within the emissive layer. The chemical structure of SiDBFCz is shown in Figure S3.

The optical and electronic properties of SiDBFCz were originally characterized in our previous work [1] and are presented here to support its selection as the host material for the Eu5NHCrown-based devices. These properties were determined by UV–Vis absorption, photoluminescence, low-temperature phosphorescence, and ultraviolet photoelectron spectroscopy (UPS).

The optical properties of SiDBFCz were measured using thermally evaporated thin films on quartz substrates. The absorption onset was used to estimate the optical gap, and the photoluminescence spectrum was recorded under 300 nm excitation.

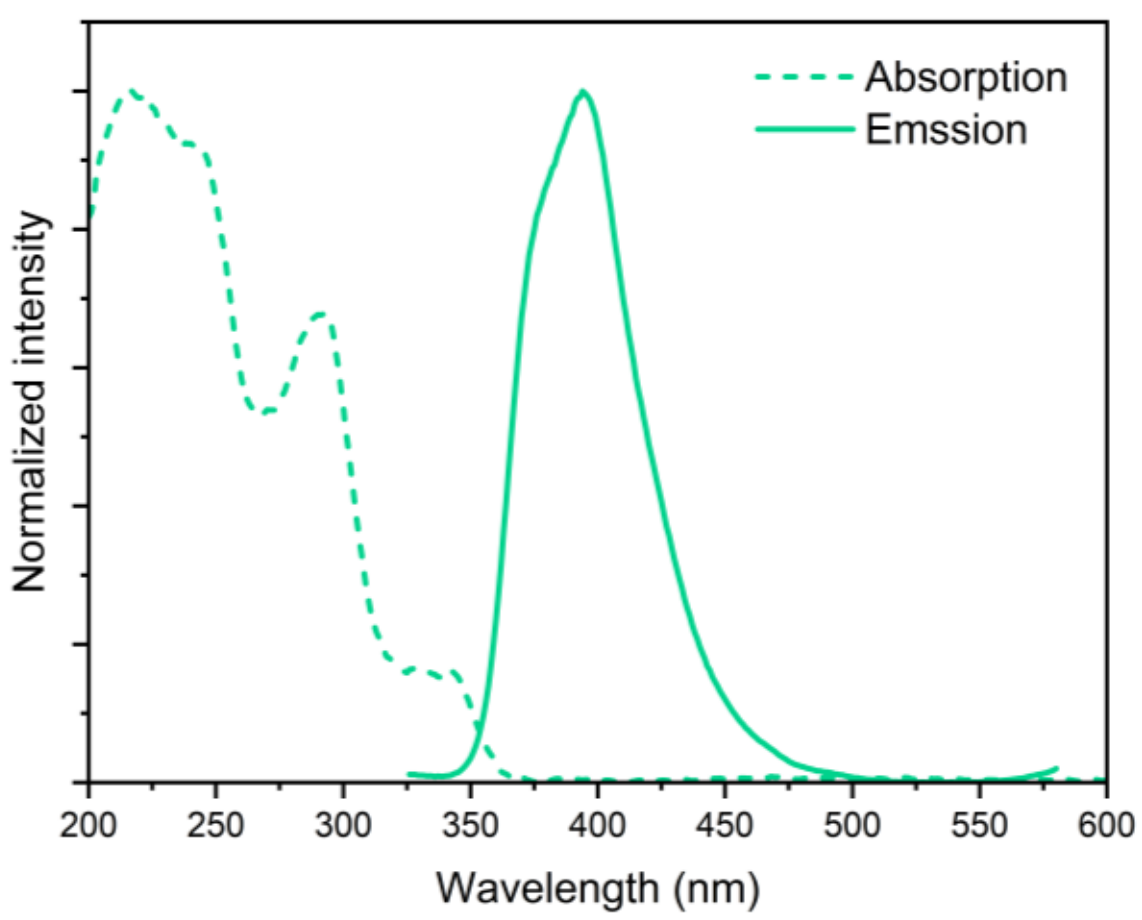

Figure S4. Absorption and emission spectra of SiDBFCz. The emission spectrum was recorded under 300 nm excitation. The absorption onset was used to determine the optical gap of SiDBFCz.

The triplet energy of SiDBFCz was determined from low-temperature phosphorescence measurements. Dilute SiDBFCz films were prepared in a PMMA matrix and measured at 77 K using 275 nm excitation. The onset of the phosphorescence spectrum was used to estimate the triplet energy (Figure S5).

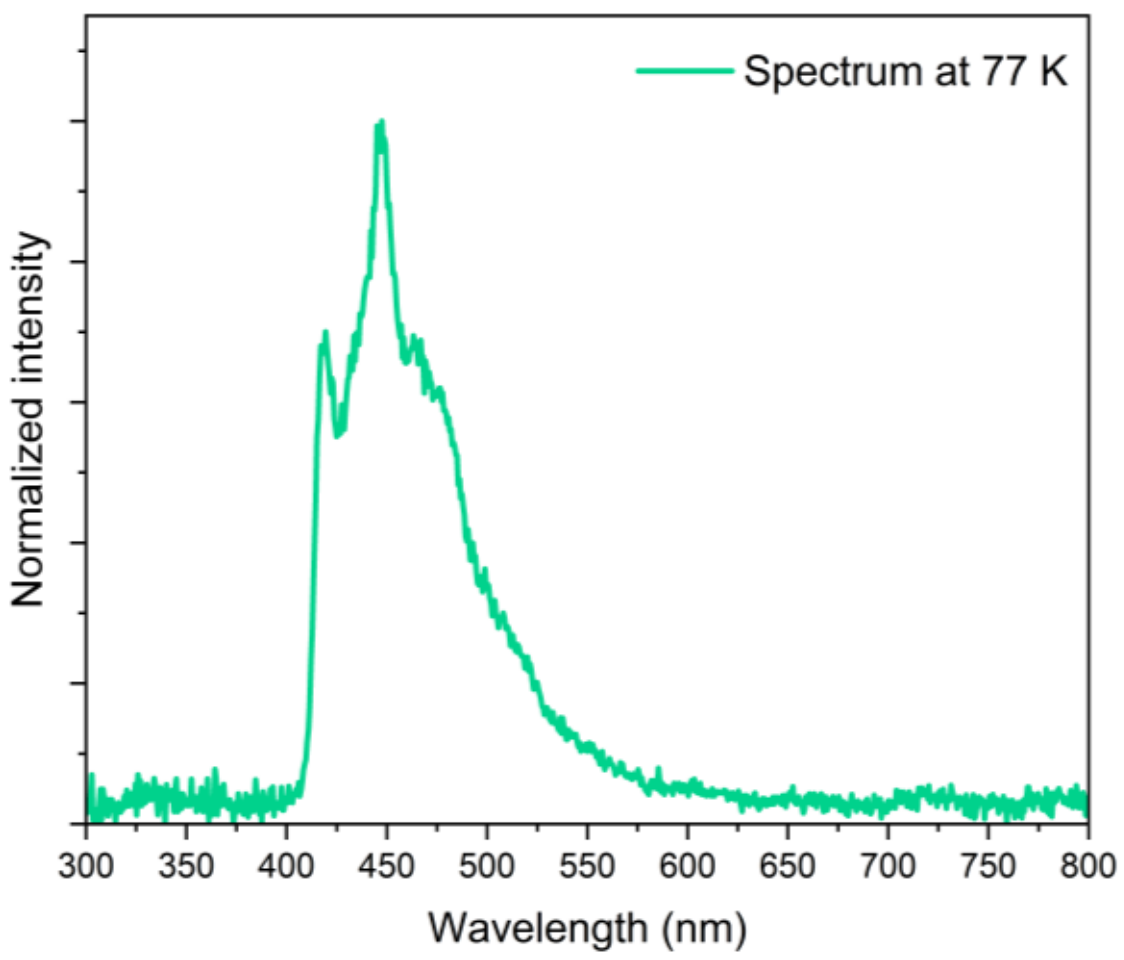

Figure S5. Low-temperature phosphorescence spectrum of SiDBFCz. Phosphorescence spectrum of SiDBFCz measured at 77 K from a dilute SiDBFCz film. The spectral onset was used to determine the triplet energy, $T_1$, of SiDBFCz.

The frontier energy levels of SiDBFCz were determined by combining ultraviolet photoelectron spectroscopy (UPS) and optical absorption measurements. Specifically, the highest occupied molecular orbital (HOMO) energy, corresponding to the ionization energy, was obtained directly from UPS, while the optical gap ($E_{opt}$) was extracted from the onset of the UV–Vis absorption spectrum. The lowest unoccupied molecular orbital (LUMO) energy was then derived from the HOMO energy and optical gap according to:

$$E_{\mathrm{LUMO}} = E_{\mathrm{HOMO}} + E_{\mathrm{opt}}, \qquad (1)$$

UPS measurements were performed using a helium discharge lamp (He I, 21.22 eV) and a hemispherical analyzer under high-vacuum conditions (~$10^{-10}$ mbar). The sample was deposited on an argon-sputtered gold substrate and transferred without exposure to ambient conditions. Energy calibration was performed using the Fermi edge of sputter-cleaned silver, yielding an instrumental resolution of 0.04 eV. Taking into account additional uncertainties arising from sample preparation, charging effects, and data evaluation, the overall measurement uncertainty is estimated to be approximately 70 meV. Ionization energies were determined from:

$$\mathrm{IE} = h\nu - E_{\mathrm{cutoff}} + E_{\mathrm{HOMO}}, \quad (2)$$

where $h\nu = 21.22$ eV, $E_{\mathrm{cutoff}}$ is the secondary electron cutoff, and $E_{\mathrm{HOMO}}$ is the HOMO onset relative to the Fermi level.

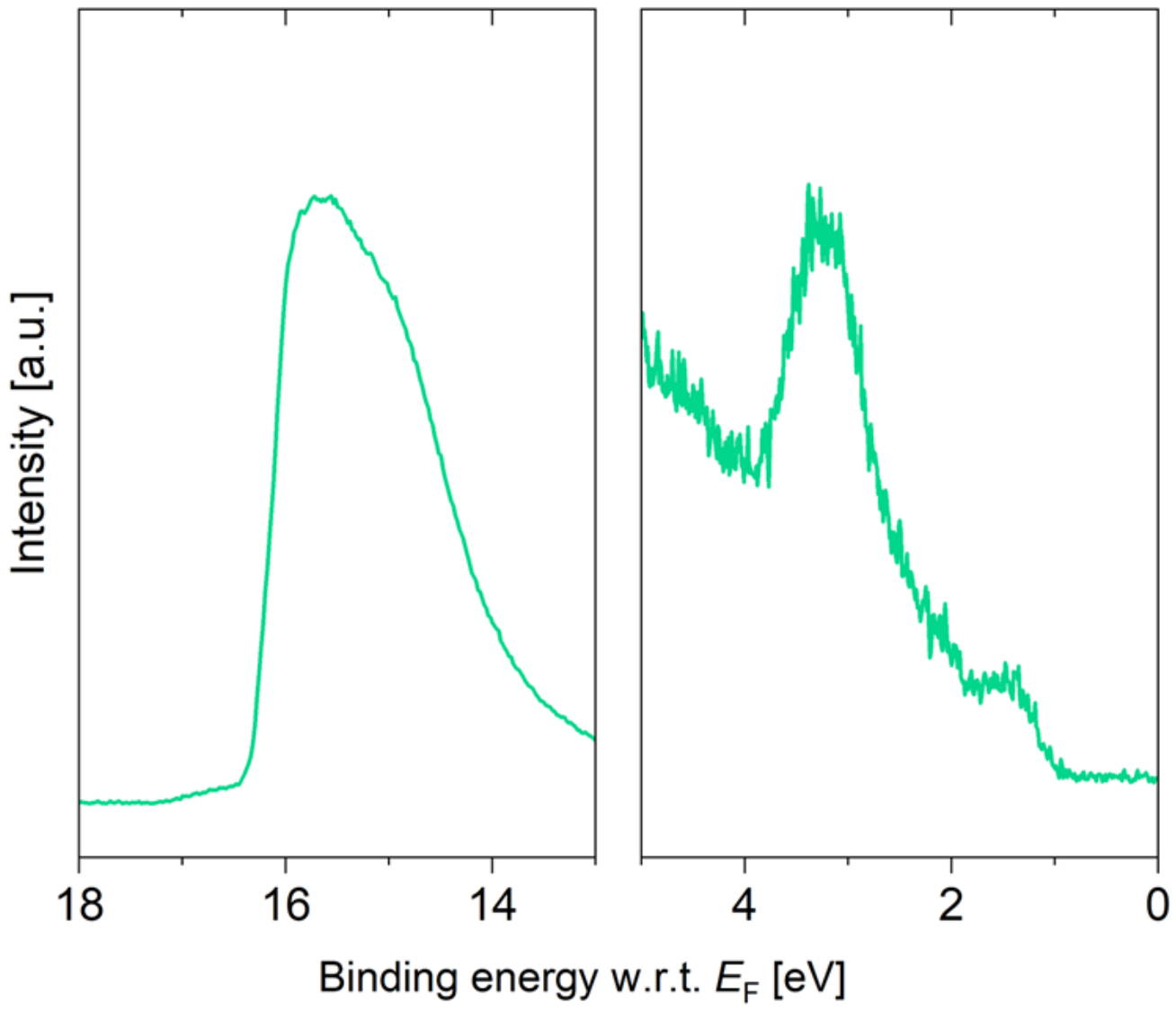

Figure S6. UPS spectra of SiDBFCz. The secondary electron cutoff and HOMO onset were used to extract the ionization energy (IE) and HOMO level.

The optical and electronic characterization of SiDBFCz is summarized in Table S1. The values include the HOMO energy obtained from UPS, the optical gap extracted from UV–Vis absorption, the LUMO energy derived from the HOMO energy and optical gap, and the triplet energy determined from low-temperature phosphorescence measurements.

Table S1. Optical and electronic characterization of SiDBFCz. All values are given in eV.

| **Host** | **HOMO** | $E_{\mathrm{opt}}$ | **LUMO** | $T_1$ |
|---|---|---|---|---|
| SiDBFCz | −6.0 | 3.4 | −2.5 | 3.0 |

### 3.2. Cavity optimization

The optical cavity of a bottom-emitting OLED strongly influences the outcoupled intensity and the apparent device efficiency.[2,3] Therefore, the thicknesses of the hole-transport and electron-transport layers were varied to identify the device architecture that provides the best optical outcoupling for the Eu5NHCrown emission. The resulting current density–voltage–luminance characteristics, EQE curves, and EL spectra are used to select the optimized device structure discussed in the main manuscript.

The cavity-optimization bottom-emitting devices were fabricated with the following structure: ITO (Indium tin oxide) / $MoO_3$ (Molybdenum trioxide) [2 nm] / SimCP2 (Bis[3,5-di(9H -carbazol-9-yl)phenyl]diphenylsilane) [50-60-70-80 nm] / SiDBFCz (9-[4-(Triphenylsilyl)-2-dibenzofuranyl]-9H-carbazole):Eu5NHCrown [30 wt%, 20 nm] / mSiTrz (2-Phenyl-4,6-bis(3-(triphenylsilyl)phenyl)-1,3,5-triazine) [5 nm] / TSPO1 (Diphenyl[4-(triphenylsilyl)phenyl]phosphine oxide):Yb (Ytterbium) [1 vol%, 30-25-20-15 nm] / Al [120 nm].

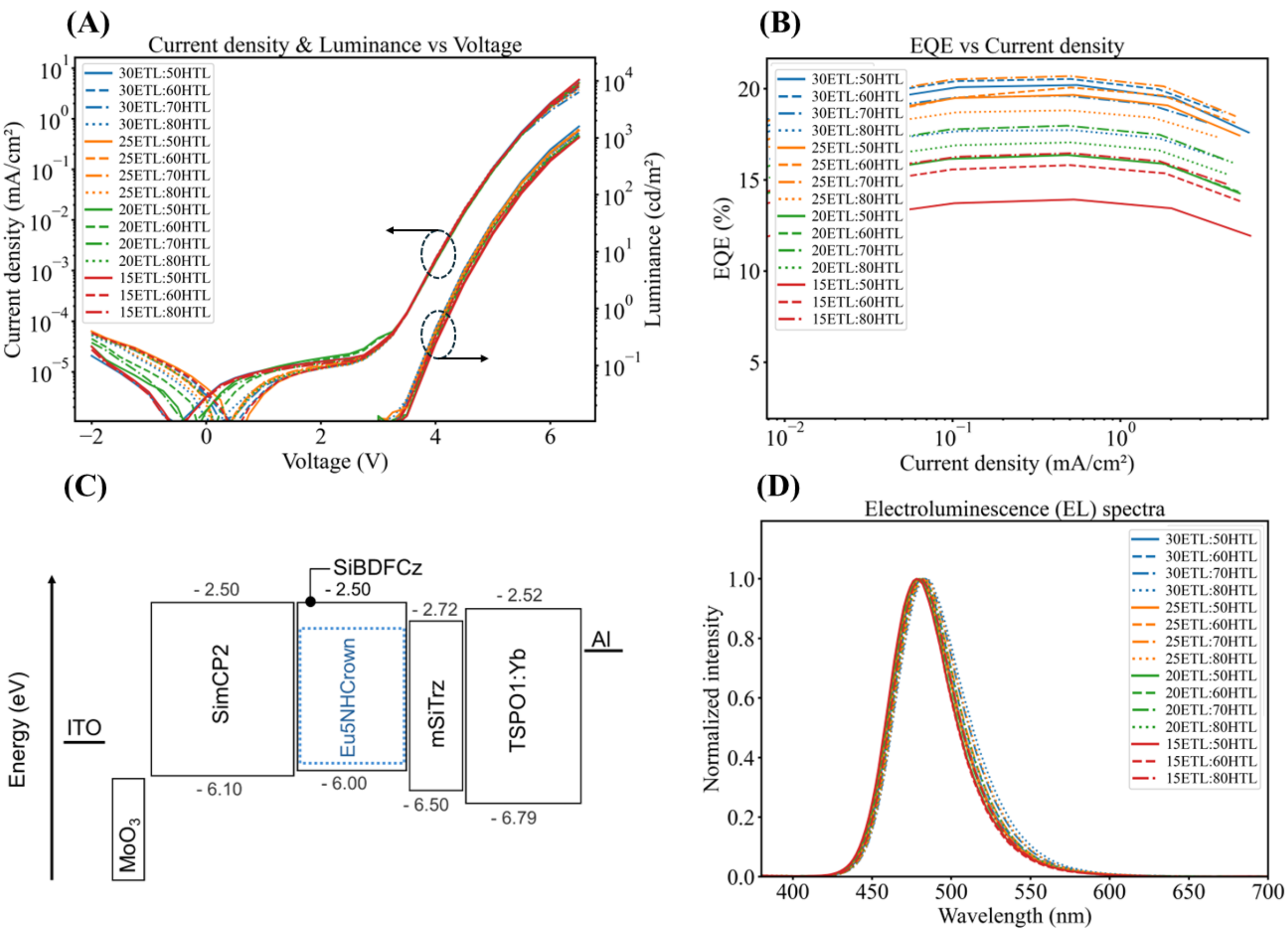

Figure S7. (A) Current density–voltage–luminance (*J–V–L*) characteristics of the devices, (B) External quantum efficiency (EQE) as a function of current density, (C) Schematic diagram of the device structure, (D) Electroluminescence (EL) spectra of the devices at 1000 cd/m$^2$.

### 3.3. Pixel statistics

To evaluate the reproducibility of the optimized Eu5NHCrown-based OLED architecture, multiple pixels from the same fabrication run were measured. The pixel statistics show that the devices exhibit highly similar and largely overlapping current density, luminance, EQE, and current-efficiency characteristics, confirming the reproducibility of the optimized device performance across different pixels.

The optimization devices were fabricated with the following structure: ITO (Indium tin oxide) / $MoO_3$ (Molybdenum trioxide) [2 nm] / SimCP2 (Bis[3,5-di(9H-carbazol-9-yl)phenyl]diphenylsilane) [70 nm] / SiDBFCz (9-[4-(Triphenylsilyl)-2-dibenzofuranyl]-9H-carbazole):Eu5NHCrown [30 wt%, 20 nm] / mSiTrz (2-Phenyl-4,6-bis(3-(triphenylsilyl)phenyl)-1,3,5-triazine) [5 nm] / TSPO1 (Diphenyl[4-(triphenylsilyl)phenyl]phosphine oxide):Yb (Ytterbium) [1 vol%, 25 nm] / Al [120 nm].

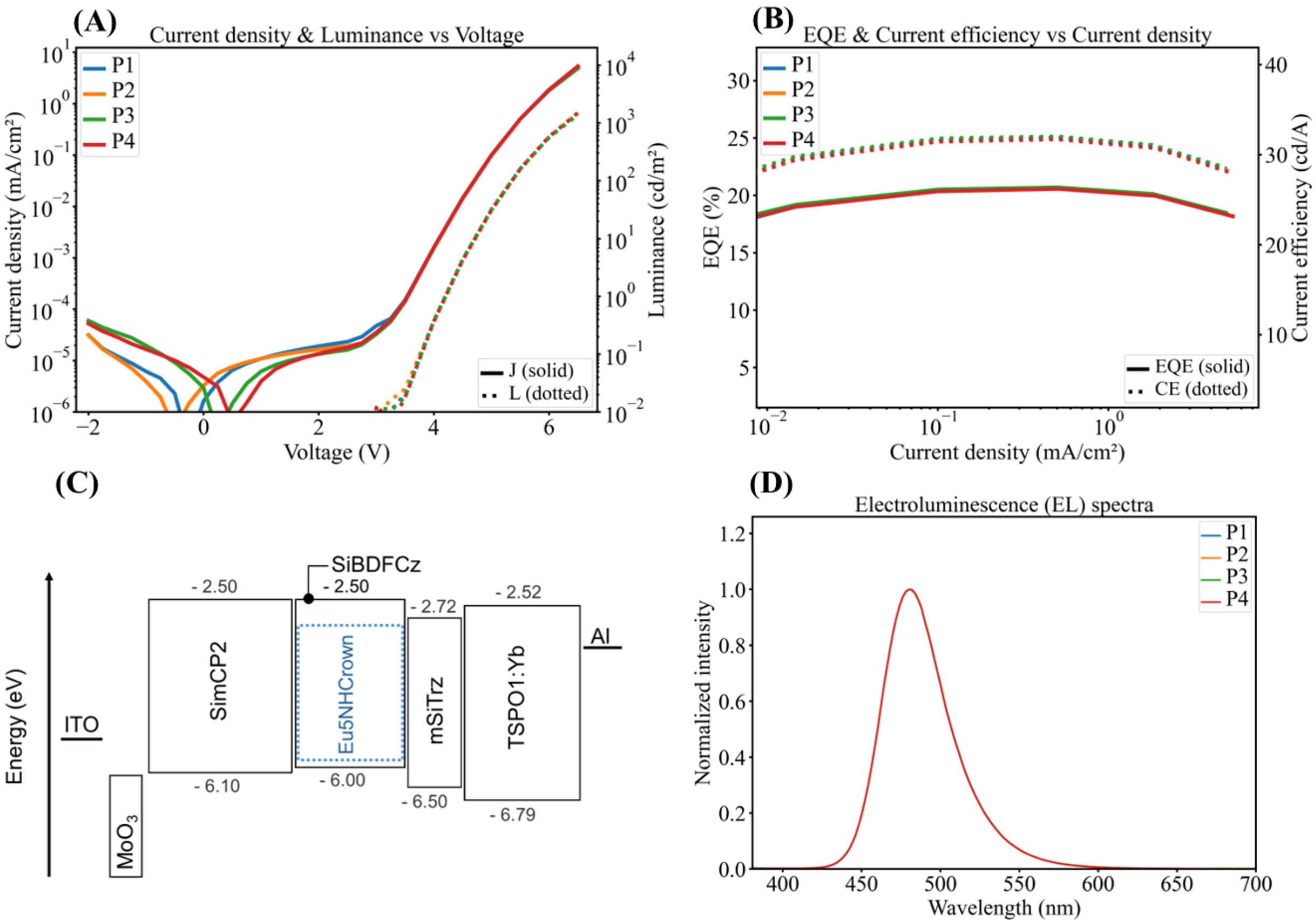

Figure S8. (A) Current density–voltage–luminance (*J*–*V*–*L*) characteristics of the devices, (B) External quantum efficiency (EQE) and current efficiency as a function of current density, (C) Schematic diagram of the device structure, (D) Electroluminescence (EL) spectra of the devices at 1000 cd/m$^2$.

### 3.4. Control devices

$Ir(ppy)_3$-based control devices were fabricated using a comparable device architecture with a well-established phosphorescent emitter. The electron-transport-layer thickness and doping concentration were adjusted to account for the different emission wavelength of $Ir(ppy)_3$, provide an appropriate optical cavity for the green control device, and optimize charge balance. The $Ir(ppy)_3$ control devices had the following structure: ITO (Indium tin oxide) / $MoO_3$ (Molybdenum trioxide) [2 nm] / SimCP2 (Bis[3,5-di(9H -carbazol-9-yl)phenyl]diphenylsilane) [70 nm] / SiDBFCz (9-[4-(Triphenylsilyl)-2-dibenzofuranyl]-9H-carbazole):$Ir(ppy)_3$ (tris[2-phenylpyridine]iridium(III)) [30 wt%, 20 nm] / mSiTrz (2-Phenyl-4,6-bis(3-(triphenylsilyl)phenyl)-1,3,5-triazine) [5 nm] / TSPO1 (Diphenyl[4-(triphenylsilyl)phenyl]phosphine oxide):Yb (Ytterbium) [1 vol%, 40 nm] / Al [120 nm].

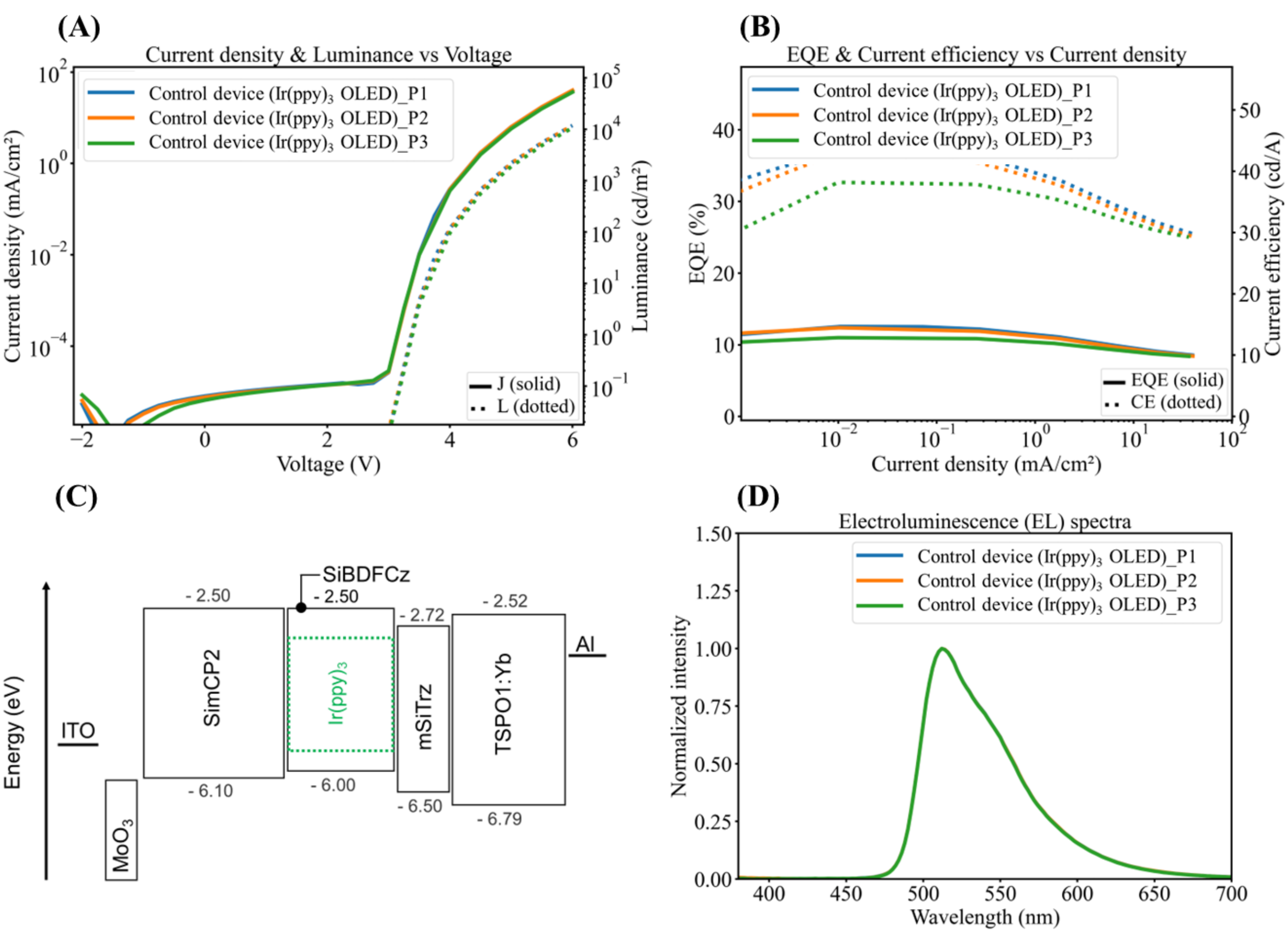

Figure S9. (A) Current density–voltage–luminance (*J*–*V*–*L*) characteristics of the control devices, (B) External quantum efficiency (EQE) and current efficiency as a function of current density, (C) Schematic diagram of the device structure, (D) Electroluminescence (EL) spectra of the devices at 1000 cd/m$^2$.

Table S2. Electroluminescent performance of the control devices.

| Device | $V_{on}$ [a] [V] | $EQE_{max}$ [b] [%] | $\eta_{max}$ [c] [cd/A] | $\lambda_{max}$ [d] [nm] | FWHM [e] [nm] |
|---|---|---|---|---|---|
| P1 | 3.4 | 12.5 | 43.5 | 512 | 60 |
| P2 | 3.4 | 12.4 | 43.0 | 512 | 60 |
| P3 | 3.4 | 11.0 | 38.0 | 512 | 60 |

(a) Turn-on voltage or voltage required to reach a luminance of 1 cd/m²; (b) Maximum External Quantum Efficiency; (c) Maximum Current Efficiency; (d) Emission Peak Wavelength; (e) Full Width at Half Maximum.

## 4. Theoretical limit of external quantum efficiency for randomly oriented emitters in planar bottom-emitting OLEDs

The external quantum efficiency (EQE) of an OLED can be expressed as

$$\eta_{\mathrm{EQE}} = \gamma \times \eta_{\mathrm{EUE}} \times q \times \eta_{\mathrm{out}}, \quad (3)$$

[2]

where $\gamma$ is the charge-balance factor, $\eta_{\mathrm{EUE}}$ is the exciton-utilization efficiency, $q$ is the intrinsic photoluminescence quantum yield, and $\eta_{\mathrm{out}}$ is the optical outcoupling efficiency. In the ideal limit where charge balance, exciton utilization, and photoluminescence quantum yield all approach unity, the EQE becomes equal to $\eta_{\mathrm{out}}$.

In a conventional planar bottom-emitting OLED (glass/ITO/organic/metal), the maximum optical outcoupling efficiency is fundamentally constrained by total internal reflection at the glass/air interface, waveguiding in the high–refractive-index organic layers and coupling to surface-plasmon modes at the metal electrode. The magnitude of these losses depends sensitively on the orientation of the emitter's transition dipole moment: emitters with predominantly horizontal dipole orientation couple less strongly to waveguided and plasmonic modes and can therefore achieve outcoupling efficiencies significantly above 30%, whereas emitters with randomly oriented dipoles distribute their emission equally among vertical and horizontal components, leading to stronger coupling into trapped optical modes and restricting the maximum outcoupling efficiency to approximately 20–22%.[2,3]

Eu(II) emitters are expected to behave as randomly oriented emitters because their electroluminescence originates from an atomic Eu(II)-centered $5d$–$4f$ transition that terminates on the $4f^7$ ($^8S_{7/2}$) ground state.[4] The $4f^7$ ($^8S_{7/2}$) ground state has zero orbital angular momentum ($L$ = 0) and an isotropic $S$-state character, so the emission is not associated with a fixed molecular transition dipole that can be preferentially oriented by molecular design.[5] Consequently, Eu(II) complexes are not expected to benefit from dipole-orientation engineering, and the random-dipole outcoupling limit of $\eta_{\mathrm{out}} \approx 20$–$22\,\%$ defines the theoretical maximum EQE in planar bottom-emitting devices.[2,3]

The Eu(II) deep-blue OLEDs reported in this work exhibit an EQE of 20.7 %. Given the isotropic nature of the Eu(II) transition dipoles and the absence of additional outcoupling structures or preferential dipole orientation, this value is effectively at the theoretical maximum for a randomly oriented emitter with near-unity internal quantum efficiency in a conventional bottom-emitting OLED stack.

## 5. Literature-reported Europium-based OLEDs

The electroluminescent performance of Eu5NHCrown is summarized alongside previously reported Eu(II)-based OLEDs in Table S3. Among the compared Eu(II)-based OLEDs, Eu5NHCrown shows the strongest overall performance, combining a high EQE of 20.7%, a low turn-on voltage of 4.2 V, blue emission at 480 nm, and a narrow EL FWHM of 44 nm. This highlights the capability of the Eu5NHCrown design to deliver blue electroluminescence with high spectral purity and near-limit external quantum efficiency in a simple OLED architecture.

Table S3. Comparison of electroluminescent performance of selected Eu(II)-based OLEDs.

| Emitter | Fabrication method | $\lambda_{EL}$ (nm) | FWHM (nm) | $V_{on}$ (V) | $EQE_{max}$ (%) | Reference |
|---|---|---|---|---|---|---|
| $Eu_2(L3)_2I_4$ | Spin coating | 485 | ≈50 | 9.6 | 9.2 | [6] |
| Eu–*tBu* | Vacuum deposition | 478 | 64 | 5.7 | 15.7 | [4] |
| $EuI_2$–$N_8$ | Vacuum deposition | 515 | ≈95 | 6.5 | 17.7 | [7] |
| Eu5NHCrown | Vacuum deposition | 480 | 44 | 4.2 | 20.7 | This work |

FWHM values marked with "≈" were estimated from the published electroluminescence spectra, as explicit EL FWHM values were not reported.